# Ferromagnetism and transport properties of the Kondo system $Ce_4Sb_{1.5}Ge_{1.5}$

V.N. Nikiforov[a], V.V. Pryadun[a], A.V. Morozkin[b], V.Yu. Irkhin*[c]

[a]*Physical Department, Moscow State University, 119899, Russia*
[b]*Chemical Department, Moscow State University, 119899, Russia*
[c]*Institute of Metal Physics, 620990 Ekaterinburg, Russia*

*E-mail: valentin.irkhin@imp.uran.ru

Ferromagnetic ordering with a small magnetic moment is found below 14 K from SQUID measurements for the compound $Ce_4Sb_{1.5}Ge_{1.5}$. The transport characteristics of a number of $Ce_4Sb_{3-x}T_x$ (T = Ge, Si, Sn, Pb, Al) systems were measured at room temperature, $Ce_4Sb_{1.5}Ge_{1.5}$ demonstrating the highest Seebeck coefficient. Its transport properties (resistivity, thermal conductivity, thermoelectric power) are experimentally investigated in detail up to low temperatures. The Kondo-lattice behavior of resistivity and its anomaly at the ordering point are found. A comparison with the Wiedemann-Franz law is performed.

**1. Introduction**

The ternary intermetallic cerium compounds form a great variety of systems with non-trivial physical properties, including Kondo lattices, intermediate valence and heavy-fermion compounds etc. In particular, they can demonstrate anomalous magnetic ordering (including small magnetic moments), which is accompanied by anomalies of electronic characteristics [1,2]. Besides that, some systems have promising thermoelectric properties, i.e. high Seebeck coefficient and figure of merit *ZT* [3].

Here, we discuss transport characteristics of a number of $Ce_4Sb_{3-x}T_x$ (T = Ge, Si, Sn, Pb, Al) alloys. As for thermoelectric properties, this system can be compared with $\beta$-$Zn_4Sb_3$ and rare-earth based anti-$Th_3P_4$ structure systems [4].

In more detail, we investigate magnetism and transport properties of the system with the composition $Ce_4Sb_{1.5}Ge_{1.5}$.which corresponds to optimal thermoelectric properties (highest Seebeck coefficient). This indicates high density of states at the Fermi level and anomalies of other electron properties.

Investigation of the related systems was performed in a number of works. The Kondo lattice $Ce_4Sb_3$ demonstrates moderately heavy-fermion behavior with $\gamma$ = 180 mJ/$K^2$ mol Ce and ferromagnetic ordering [5]. According to [5], a very sharp peak of specific heat at 3.9 K in zero magnetic field corresponds to the phase transition from paramagnetic to ferromagnetic order. Specific heat data indicate that antiferromagnetic components may exist in the ferromagnetic state, as well as in other $R_4X_3$ intermetallics [6]. The presence of such a complicated magnetism has been further confirmed by the neutron diffraction study that evidences commensurate antiferromagnetic ordering at 2 K in zero magnetic field. Saturation magnetization value of only 0.93 $\mu_B$/Ce ion was observed at 1.8 K [7].

Magnetic properties of Ce-Co-Ge system were discussed in Refs.[8,9]. Some phases (including $Ce_4Ge_3$) which exhibit ferromagnetic-like ordering (or complex ordering including a ferromagnetic component) with low Curie temperature were also considered.

The magnetic structures of $Ce_4Sb_3$ [7] and $Ce_4Ge_3$ [6] turn out to be not compatible from the point of view of their magnetic symmetry: they demonstrate commensurate antiferromagnetic order with different orderings of $Ce_4$ cluster. Therefore, the solid solution $Ce_4Sb_{1.5}Ge_{1.5}$ is an interesting object for physical investigation. In the present paper, we find in $Ce_4Sb_{1.5}Ge_{1.5}$ ferromagnetism with strongly reduced moments, which is characteristic for the Kondo systems.

## 2. Experimental

The polycrystalline samples of $Ce_4Sb_{1.5}Ge_{1.5}$ were synthesized by melting the starting mixture followed by annealing. The preparation technique included an electric arc furnace under an argon atmosphere using a non-consumable tungsten electrode and a water-cooled copper tray [10]. The purity of the component metals was better than 99%. The quality of the samples before physical measurements was determined using X-ray phase analysis and microprobe X-ray analysis. X-ray data were obtained on a diffractometer DRON-3.0 (Cu K radiation, 2 $\Theta$ = 20–70°, step 0.05°, 5 s per step). The diffractograms obtained were identified by means of calculated patterns using the Rietan-program in the isotropic approximation. A Camebax Micro-analyzer was employed to perform microprobe X-ray spectral analyses and SEM images of the samples [11].

The $Ce_4Sb_{1.5}Ge_{1.5}$ system has anti-$Th_3P_4$ cubic structure. For the sample investigated, XRD yields practically single phase (about 2 mass % of CeSb impurity). Lattice parameter is $a$ = 0.93807(9) nm, $X_{Ce}$ =0.5672(4), RF = 6.5%.

We synthesized and investigated also a large number of Ce compounds and alloys containing atoms of simple and transition elements. In our case, X-ray structure analysis enabled us to determine the lattice parameters. These data are compiled in the Table 1.

Table 1.
Physical properties at $T$ =300 K of $Ce_4Sb_{3-x}T_x$ systems and related compounds: electric resistivity ρ (μOhm m), Seebeck coefficient $S$, thermal conductivity κ and figure of merit parameter $ZT = S^2T/\rho \kappa$. The Wiedemann-Franz parameter is WF = ρ κ/$L_0 T$, $L_0 = (\pi^2/3) (k_B/e)^2$ being the free electron Lorentz number ($k_B$ is the Boltzmann constant and $e$ the electronic charge). The data on $Zn_4Sb_3$ are presented in comparison with Ref. [12]. The lattice parameter $a$ was investigated not for all systems

| Compound | $a$, nm | $S$, μV/K | ρ, μΩ m | κ, W/m K | WF | $ZT$ |
|---|---|---|---|---|---|---|
| $Zn_4Sb_3$ [12] | | 114.3 | 20.42 | | | |
| $Zn_4Sb_3$ | | 128 | 29.0 | 5.9 | 23.3 | 0.029 |
| $Ce_4Sb_3$ | 0.95302(5) | 8.7 | 1.69 | 8,96 | 2.06 | 0.0015 |
| $Ce_4Sb_{2.2}Si_{0.8}$ | 0.94470(24) | 13.0 | 3.2 | 8.2 | 3.58 | 0.0019 |
| $Ce_4Sb_2Si$ | 0.94326(32) | 13.7 | 2.1 | 8.0 | 2.29 | 0.0034 |
| $Ce_4Sb_{1.5}Si_{1.5}$ | 0.93395(25) | 11.4 | 13.8 | - | - | - |
| $Ce_4Sb_{2.2}Ge_{0.8}$ | 0.94454(12) | 13.7 | 1.97 | 8.0 | 2.15 | 0.0036 |
| $Ce_4Sb_2Ge$ | 0.94367(15) | 13.7 | 2.1 | 8.0 | 2.29 | 0.0034 |
| $Ce_4Sb_{1.5}Ge_{1.5}$ | 0.93807(9) | 17.6 | 2.15 | 5.7 | 1.7 | 0,0076 |
| $Ce_4Ge_3$ | 0.92173(10) | 1.7 | 17.2 | 5.1 | 11.93 | 0,00001 |
| $Ce_4Sb_2Sn$ | | 12.8 | 2.42 | 7.9 | 2.60 | 0.0026 |
| $Ce_4Sb_2Sn_{0.5}Pb_{0.5}$ | | 9.98 | 1.54 | 9.5 | 2.0 | 0.0020 |
| $Ce_{58}Sb_{36}Pb_6$ | 0.95345(13) | 8.2 | 1.82 | 9.4 | 2.33 | 0.0012 |
| $Ce_{57}Sb_{42}Al_1$ | | 2.6 | 2.15 | 7.3 | 2.14 | 0.00013 |
| $Ce_{57}Sb_{42}Ga_1$ | | 6.1 | 1.77 | 8.1 | 1.95 | 0.00078 |
| $Ce_{59}Sb_{39}In_2$ | | 5.6 | 1.46 | 9.4 | 1.87 | 0.00069 |
| $La_{58}Sb_{22}Pb_{20}$ | | 17.0 | 2.58 | 7.5 | 2.63 | 0.0045 |
| $Pr_4Sb_2Si$ | | 11.0 | 1.5 | 8.2 | 1.67 | 0.0030 |

The transport measurements were carried out in the temperature range of 2.0–380K by the Quantum Design Physical Property Measurement System. This enables simultaneous measurements of thermal conductivity, Seebeck coefficient and resistivity by monitoring both the temperature and voltage drop across a sample as a heat pulse is applied to one end. Measurements were performed under high vacuum (~$10^{-4}$ torr) using a four-probe lead configuration. The software of this system uses adaptive algorithms to optimize measurement parameters such as heater current, heat pulse period and has a correcting for radiative heat losses and correcting for Seebeck coefficient of manganin leads. The detailed information on the measuring technique is presented in Ref. [13].

Electrical resistance along the long axis of samples was measured employing standard four-contact geometry at a constant current of 0.8 mA (we found no overheating of samples at this magnitude of current).

The Seebeck coefficient (thermopower at zero electric current) of the samples was measured when they were disconnected from the current source. The heat flow in the samples was produced by an electrical heater. The thermally induced voltage $E$ was measured through the same potential contacts used in measuring resistances. The temperature difference $dT$ between these contacts was measured by a differential manganin–constantan thermocouple that was cemented to copper leads. This thermocouple was fabricated from thin wires of manganin (7 μm) and constantan (30 μm). The Seebeck coefficient was calculated as the ratio $E/dT$. Uncertainty in the $S$ value was estimated to be smaller than 10%.

The thermal conductivity was measured by the method of the longitudinal stable state. The temperature difference across the samples was produced by the electrical heater. The sample was clamped to the other end of the copper base. Heat losses from the sample were minimized by evacuating the chamber. The temperature of the copper base was measured using a chromel–alumel thermocouple. The temperature difference across the samples was measured by the differential manganin–constantan thermocouple. The random error of thermal conductivity was smaller than 3%.

The magnetization was measured by a SQUID magnetometer of Quantum Design MPMS-5. The temperature dependence of magnetic moment was also experimentally measured by the SQUID magnetometer [14].

## 3. Magnetic properties

We studied the field and temperature dependences of the magnetization for $Ce_4Sb_{1.5}Ge_{1.5}$ in the broad interval of magnetic fields ($H < 50$ kOe) and temperatures ($2 < T < 300$ K). The measurements demonstrated occurrence of a small ferromagnetic moment below about 13 K.

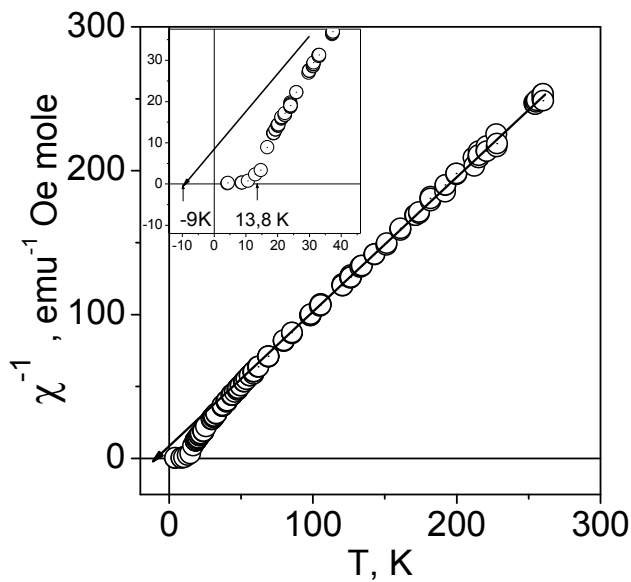

Fig.1 Temperature dependence of inverse magnetic susceptibility from SQUID measurements. The inset shows low-temperature behavior

Despite divergence of magnetic susceptibility near 13 K, the paramagnetic Curie temperature $\theta = -9$ K is negative (Fig.1), which is typical for the Kondo systems where on-site antiferromagnetic Kondo coupling dominates over intersite magnetic interactions [1]. The small value of saturation moment obtained (about $0.1\mu_B$, Fig.2) is also typical for the dense Kondo systems because of the Kondo screening. A similar situation takes place, e.g., in the compound $CeRuSi_2$ [2].

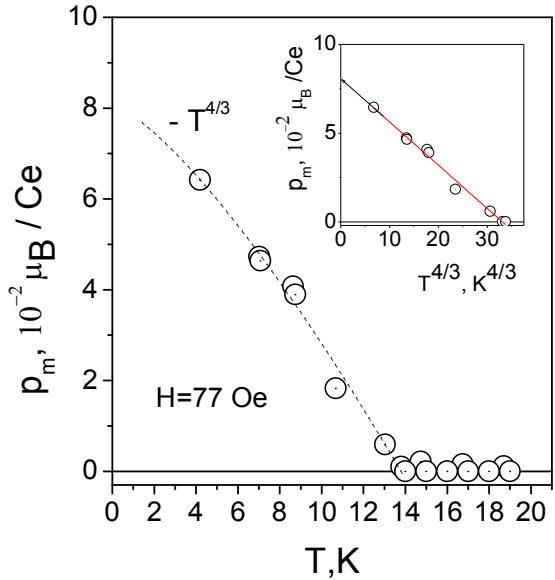

Fig.2. Temperature dependence of static magnetic moment from SQUID measurements. The inset shows the $T^{4/3}$–interpolation

The $T^{4/3}$–interpolation characteristic for weak itinerant ferromagnetism with strong non-spin-wave fluctuations turns out to fit our data somewhat better than the $T^{3/2}$–interpolation (spin-wave theory for localized systems). This also confirms unusual nature of magnetic state.

### 4. Transport properties

The transport characteristics of a number of $Ce_4Sb_{3-x}T_x$ (T = Ge, Si, Sn, Pb, Al) systems measured at room temperature are presented in Table 1. One can see that $Ce_4Sb_{1.5}Ge_{1.5}$ has largest value of the Seebeck coefficient. Therefore we present a detailed investigation of its properties in a wide temperature region.

The $Ce_4Sb_{1.5}Si_{1.5}$ sample has rather large residual resistivity about $1.4 \cdot 10^{-7}$ Ohm m. The temperature dependence of resistivity at low temperatures is shown in Fig.3. A nearly linear behavior is observed below magnetic transition temperature (about 13 K) where inflection point of $\rho(T)$ occurs. This is connected with that magnetic scattering becomes weakened in the ordered phase. Such a behavior is similar to that observed in ferromagnetic $CeRuSi_2$ [2,15].

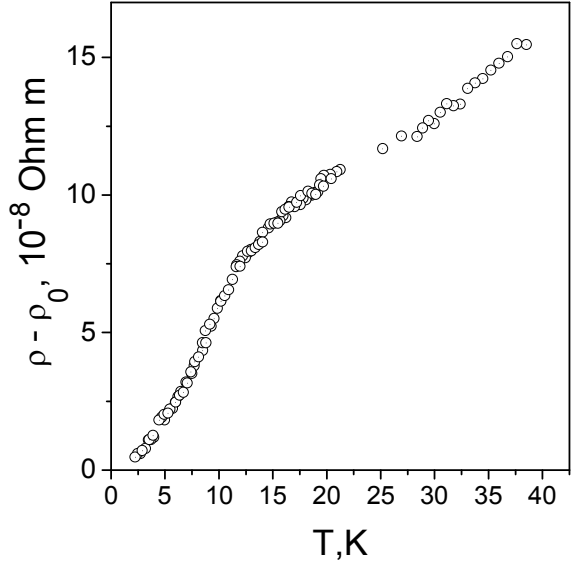

Fig. 3. The temperature dependence of resistivity at low temperatures

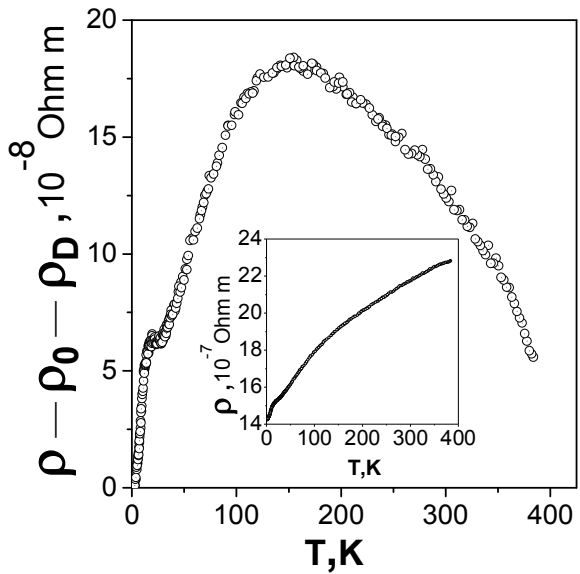

Fig.4. The temperature dependence of resistivity in a wide temperature region with residual and lattice resistivity, $\rho_0$ and $\rho_D$, being subtracted. The inset shows the total resistivity

From the data of Ref. [5] on $\beta T^3$-term in specific heat of $Ce_4Sb_3$, we can estimate the Debye temperature as $\theta_D = 200$ K. Then, by using the Bloch-Grüneisen formula we can extract the lattice (phonon) contribution to resistivity (Fig.4). Thus magnetic resistivity demonstrates a high-temperature maximum near 150 K, which is characteristic for Kondo systems and can be approximately described by a logarithmic law in some temperature interval (cf. also the data on $CeRuSi_2$ [2]).

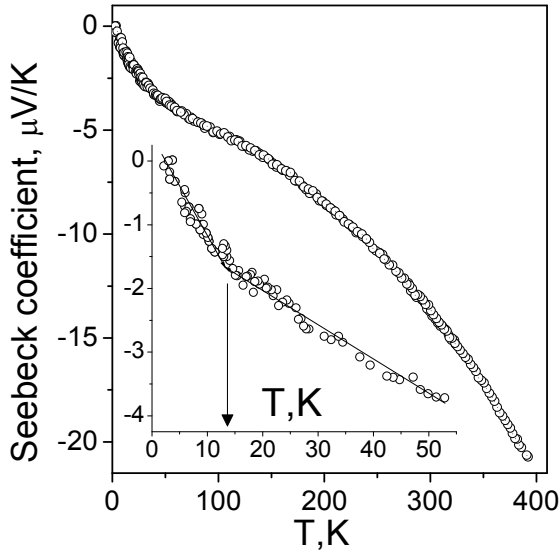

Fig.5. The temperature dependence of Seebeck coefficient. The inset shows the behavior at low temperatures (in particular, near the ordering point)

The temperature dependence of thermoelectric power is shown in Fig.5. One can see that, as well as in resistivity, an inflection point seems to occur near the Curie temperature.

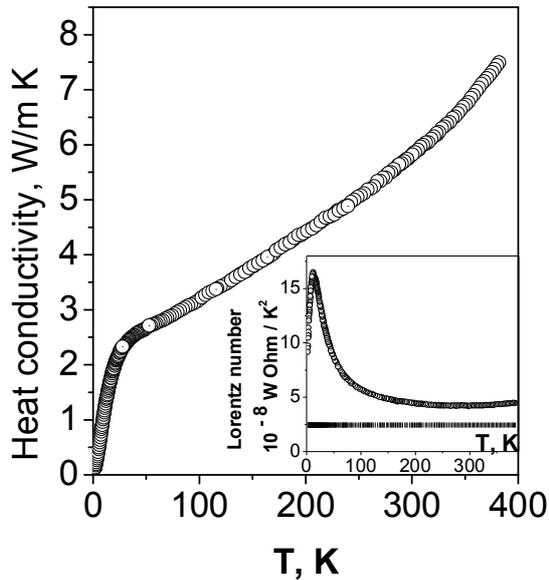

Fig.6. The temperature dependence of thermal conductivity $\kappa$. The inset shows the temperature dependence of the modified Lorentz number $L = \kappa\rho/T$; horizontal line corresponds to the Wiedemann-Franz law where $L_0 = (\pi^2/3)(k_B/e)^2$

Data on thermal conductivity are presented in Fig.6. together with the temperature dependence of the modified Lorentz number $L = \kappa\rho/T$. A comparison with the Wiedemann-Franz demonstrates that electron and phonon contributions are comparable, and the latter contribution is not dominating one at high temperature. Nevertheless, despite large value of the Seebeck coefficient, the figure of merit $ZT$ reaches at room temperatures about 0.8% only because of rather high thermal conductivity (see Table 1).

## 5. Discussion and conclusions

The results obtained demonstrate that $Ce_4Sb_{1.5}Ge_{1.5}$ is a ferromagnetic Kondo lattice. Although magnetic ordering is a rather usual phenomenon in Kondo systems, examples of ferromagnetic Kondo compounds are quite not numerous [1,2]. The presence of magnetic transition is confirmed by direct SQUID measurements and the presence of anomalies in transport properties (resistivity and thermoelectric power).

Our results on transport characteristics can be compared with typical dependences of resistivity and Seebeck coefficient including the Kondo features, which are presented in Ref. [9]. The magnetic phase transitions in intermetallic compounds usually manifest themselves in the thermoelectric power as inflections points in its temperature variation, and frequently are not discernible at all. In particular, the ferromagnetic orderings are hardly visible in thermoelectric properties of a number of $Ce_nCoGe_m$, systems. A similar situation takes place for the resistivity.

**Acknowledgements**
This work is supported in part by the Programs of fundamental research of the Ural Branch of RAS "Quantum macrophysics and nonlinear dynamics", project No. 12-T-2-1001 and of RAS Presidium "Quantum mesoscopic and disordered structures", project No. 12-P-2-1041.